\documentclass[12pt, preprint]{emulateapj}
\usepackage{graphicx}
\usepackage{epstopdf}
\shorttitle{Contribution to the Extragalactic Gamma-ray Background}
\shortauthors{Venters}

\usepackage{epsf}
\begin{document}

\title{Contribution to the Extragalactic Gamma-ray Background from the Cascades of Very-high Energy Gamma Rays from Blazars}

\author{Tonia M. Venters\altaffilmark{1,2}}
\affil{Department of Astronomy and Astrophysics, The University of Chicago, Chicago, IL 60637, USA}
\email{moira@uchicago.edu}

\altaffiltext{1}{Presented as part of a dissertation to the Department of Astronomy and Astrophysics, The University of Chicago, in partial fulfillment of the requirements for the Ph.D. degree.}
\altaffiltext{2}{NASA Postdoctoral Program Fellow, Goddard Space Flight Center, Greenbelt, MD 20771}

\begin{abstract}
As very high energy (VHE) photons propagate through the extragalactic background light (EBL), they interact with the soft photons and initiate electromagnetic cascades of lower energy photons and electrons. The collective intensity of a cosmological population emitting at VHEs will be attenuated at the highest energies through interactions with the EBL and enhanced at lower energies by the resulting cascade. We calculate the cascade radiation created by VHE photons produced by blazars and investigate the effects of cascades on the collective intensity of blazars and the resulting effects on the extragalactic gamma-ray background. We find that cascade radiation greatly enhances the collective intensity from blazars at high energies before turning over due to attenuation. The prominence of the resulting features depends on the blazar gamma-ray luminosity function, spectral index distribution, and the model of the EBL. We additionally calculate the cascade radiation from the distinct spectral sub-populations of blazars, BL Lacertae (BL Lacs) objects and flat-spectrum radio quasars (FSRQs), finding that the collective intensity of BL Lacs is considerably more enhanced by cascade radiation than that of the FSRQs. Finally, we discuss the implications that this analysis and upcoming \emph{Fermi} observations could have for the nature of the EBL, the evolution of blazars, blazar spectra, and other sources of gamma-ray radiation.
\end{abstract}


\maketitle

\section{Introduction}
\label{cha:intro}

The gamma-ray sky consists of resolved point sources (such as normal galaxies and active galaxies, pulsars, etc.), transient gamma-ray sources (e.g., gamma-ray bursts), and the diffuse gamma-ray radiation comprised of emission from the Galaxy and the extragalactic gamma-ray background (EGRB). 
The origins of the EGRB are, as yet, unknown; however, it is expected that the intrinsic emission of unresolved, extragalactic point sources comprises a sizable contribution to the EGRB. Additionally, many of these extragalactic point sources are also sources of very high energy (VHE) gamma rays\footnote{In this paper, we take VHE to be $\sim$ TeV.}, which interact with the soft photons of the extragalactic background light (EBL) to create electromagnetic (EM) cascades, giving rise to another contribution to the EGRB.

The gamma-ray sky, in general, was first observed by the SAS-2 and COS-B satellites in the 1970s. During the 1990s, the energetic gamma-ray experiment telescope (EGRET) aboard the {\it Compton Gamma-ray Observatory} observed the gamma-ray sky and, in particular, resolved 271 gamma-ray sources (the third catalog of EGRET point sources; \citealt{har99}) of which 93 were identified, either confidently or potentially, as blazars (gamma-ray loud active galactic nuclei (AGNs)). Currently, the \emph{Fermi Gamma-ray Space Telescope} is taking data and in its first few months of observations has already identified 109 blazars (LAT Bright AGN Sample (LBAS); \citealt{lat09a}). Thus, blazars constitute the class of gamma-ray emitters with the largest number of identified members.

Blazars, specifically, are those AGNs for which the jet is closely aligned
with the observer's line of sight \citep{bla79}, and include both BL Lac objects
and gamma-ray-loud flat spectrum radio quasars (FSRQs). 
Since resolved blazars comprise the class of gamma-ray emitters with the largest number of identified members, \emph{unresolved} blazars are expected to have a sizable contribution to the EGRB \citep{sre98}.  However, the exact amount of this contribution is, as yet, unclear due to the uncertainty surrounding the distribution of blazars with respect to luminosity and redshift, the blazar gamma-ray luminosity function \citep[GLF;][hereafter NT06]{pad93,ste93,sal94,chi95,ss96a,kaz97,chi98,muk99,muc00,der07,gio06,nt06}. 



In addition to the dependence on the blazar GLF, the blazar contribution to the EGRB also depends on the nature of blazar spectra at GeV energies. The energy spectrum of the gamma-ray emission of blazars at these energies has traditionally been approximated by a power law, defined by a spectral index, which, in most cases, has a value between $1.5$ and $3$ for individual blazars.  The distribution of blazar gamma-ray spectral indices, the spectral index distribution (SID), is a critical input in the determination of the shape of the spectrum of the collective unresolved blazar intensity that is the blazar contribution to the EGRB \citep{ss96a,poh97,pv08,ven09}. The spread in the blazar SID determines the fraction of blazars with hard spectra, which will contribute most significantly at high energies and hence will introduce curvature in the shape of the unresolved blazar intensity \citep{ss96a,vp07,pv08,ven09}. Since it is the highest energy photons that interact with EBL photons, the determination of the blazar SID is crucial to our understanding of the contribution of EM cascade radiation to the EGRB.

The EBL is composed of emission from starlight (at optical, ultraviolet, and near-infrared wavelengths) and reradiated thermal dust emission (at far-infrared) in galaxies. At observed energies beyond the EGRET energy range (but well within the \emph{Fermi} energy range), photons suffer significant attenuation due to pair production interactions with the soft photons of the EBL \citep{sal98,chen04,kne02,kne04,kne07,ste06,ste07,pri08,gil09,ven09}. In \citet{ven09}, we demonstrated that blazars 
exhibit an absorption feature at the highest energies in their collective spectrum. We showed that the strength of such an absorption feature depends on the blazar GLF with the strongest feature arising for those models that situate more high-luminosity sources at high redshifts. We also showed that the shape of the absorption feature depends on the EBL model. Thus, we demonstrated that if blazars dominate the EGRB, the measurement of the absorption feature at the highest energies can place constraints on the blazar GLF and the EBL.  We did not, however, propagate high energy photons to determine the resulting cascade radiation, which will further alter the shape of the collective unresolved blazar spectrum. In this paper, we study the effect of the cascades resulting from propagation on the collective blazar intensity.

In interacting with EBL photons, VHE gamma rays will produce pairs of electrons and positrons, which will inverse Compton scatter EBL photons to high energies. These upscattered photons will, in turn, interact with soft EBL photons through pair production, and this ``EM cascade'' process continues until the energies of the resulting photons are low enough that pair production is no longer efficient. For any cosmological population emitting gamma rays at VHEs, the effect of EM cascading results in a flux suppression at the highest energies and enhancement at lower energies.  In the case of blazars, which can comprise a sizable contribution to the EGRB, predictions of the resulting enhancement at lower energies can lead to the overproduction of the EGRB if the collective high-energy intensity of blazars is high and\slash or the EBL is high \citep{cop97}. More recently, \citet{kne08} and \citet{ino08} have estimated the contribution of cascade radiation to the EGRB.  They insightfully demonstrate that in including radiation from EM cascades, blazars can account for nearly all of the EGRB ($\sim 80\%$-$90$\%).  However, in neither analysis was the blazar SID as determined from either EGRET or \emph{Fermi} data included, and thus, the impact of the population of blazar GeV spectral indices was not fully examined.  Furthermore, neither analysis makes use of a Monte Carlo propagation scheme, and hence do not fairly sample secondary electron energies.

In calculating the spectrum of cascade photons, we make use of a Monte Carlo propagation code called {\it Cascata}, which samples secondary electron energies from cross-section-weighted distributions (see Section \ref{subsec-propcode} and the Appendix). Once sampled, the resulting spectrum of EM cascade radiation is then calculated. In this manner, {\it Cascata} calculates a more fully sampled cascade spectrum resulting from a spectrum of primary photons.

In this paper, we revisit the contribution of EM cascade radiation to the EGRB.  Specifically, we study the impact of the blazar GLF, the blazar SID, and the EBL model on the spectrum of cascade radiation, which, in turn, affects the blazar contribution to the EGRB.  All of the aforementioned inputs remain quite uncertain.  However, if blazars do, in fact, comprise the bulk of the EGRB intensity, then the careful study of EM cascades along with \emph{Fermi} observations can be used to constrain the inputs of the collective intensity and the nature of the EBL.  Furthermore, as the EM cascade radiation is sensitive to the blazar SID, it is also sensitive to possible breaks in blazar spectra beyond the EGRET energy range (e.g., spectral breaks or cutoffs).  Thus, the study of EM cascades together with observations of the EGRB can also provide information about blazar spectra beyond tens of GeV, which can, in turn, provide insight into the nature of blazar emission.

In this paper, we demonstrate that with a Monte Carlo propagation code such as {\it Cascata} and independent constraints on the inputs from \emph{Fermi} observations, as well as those from imaging atmospheric Cherenkov telescopes (IACTs), the blazar contribution to the EGRB will ultimately be determined. In Section 2, we present the formalism of the calculation of the collective unresolved blazar intensity and a short discussion of the aspects of the code relevant to the propagation of VHE gamma rays. In Section 3, we discuss the inputs of the calculation and their uncertainties. In Section 4, we present the results of the calculation, and we discuss these results in Section 5.

\section{Formalism}
\label{cha:formalism}

\subsection{Contribution to the EGRB from Cascades of VHE Photons from Blazars}\label{subsec-contrcas}

The contribution to the EGRB due to blazars can be viewed as the superposition of the collective intensity of \emph{intrinsic} blazar spectra and the contribution from the cascade radiation from the interactions of VHE photons from blazars with the EBL:
\begin{equation}
I_E(E_0) = I^i_E(E_0) + I^c_E(E_0)\,,
\end{equation}
where the intensity, $I_E(E_0)$, is given in units of photons per unit energy per unit time per unit area per unit solid angle emitted at observer frame energy, $E_0$. The collective intensity of \emph{intrinsic} blazar spectra including attenuation by the EBL is given by \citep[for derivation, see ][]{ven09}:
\begin{widetext}
\begin{equation}
I^i_E(E_0)=\!\! \frac{c}{H_0} \frac{1}{4\pi E_f^2} \! \int_{-\infty}^{\infty} \!\!\!\! d\alpha\,p_L(\alpha)\!\!\left(\frac{E_0}{E_f}\right)^{-\alpha}\!\!\!\! \int_{z=0}^{z_{\rm max}} \!\!\!\!\!\! dz'\,\frac{e^{-\tau(E_0,z')}}{(1+z')^{\alpha}[\Omega_{\Lambda}+\Omega_m(1+z')^3]^{1/2}} \!\! \int_{L_{\gamma,{\rm min}}}^{L_{\gamma,{\rm max}}} \!\!\!\!\!\! dL_{\gamma}L_{\gamma}\rho_{\gamma},
\end{equation}
\end{widetext}
where $E_f$ is some fiducial \emph{observer} frame energy (taken to be $100$ MeV), \linebreak $L_{\gamma} = E_{f}'^2dN(E_{f}')/dtdE$ is the isotropic gamma-ray luminosity of a blazar at the fiducial \emph{rest} frame energy, $E_f' = E_f(1+z')$, $\rho_{\gamma} = d^2N/dL_{\gamma}dV_{\rm com}$ is the blazar GLF, $\tau(E_0,z)$ is the optical depth as a function of \emph{observed} photon energy and source redshift (see Section \ref{subsec-EBL}), $p_L(\alpha) = dN/d\alpha$ is the SID of gamma-ray blazars corrected for measurement error and sample bias (see Section \ref{subsec-blsp}), and $L_{\gamma,{\rm max}} = 4\pi d_L^2 (\alpha-1)(1+z)^{\alpha-2} E_f F_{\gamma,{\rm min}}$, where $F_{\gamma,{\rm min}}$ is the sensitivity of the survey above the fiducial observer frame energy\footnote{For EGRET, the sensitivity above $100$ MeV is $10^{-7}$ $\mbox{photons } {\rm cm}^{-2} {\rm s}^{-1}$. For  \emph{Fermi} LAT, the one-year sensitivity above $100$ MeV is $2 \times 10^{-9}$ $\mbox{photons } {\rm cm}^{-2} {\rm s}^{-1}$.}.  In deriving the above equation, we have assumed the standard $\Lambda$CDM cosmology and that blazar energy spectra can be described as single power laws defined by the spectral index, $\alpha$ ($dN_{\gamma}/dE_{\gamma} \propto E_{\gamma}^{-\alpha}$).

Making the dependencies explicit, the cascade intensity is given by
\begin{equation}\label{eqn-cascadeintens}
I^c_E = \frac{d^4N^c}{dtdAd\Omega dE}\,,
\end{equation}
where $dN^c/dE$ is the spectrum of cascade photons due to pair production and Inverse Compton scattering (for derivation, see the Appendix):
\begin{widetext}
\begin{equation}\label{eqn-cascadespec}
\frac{dN^c}{dE_0}(E_0) = \! \int_0^{z_{\rm max}} \!\!\! \int_{E_{p,{\rm min}}}^{E_{p,{\rm max}}} (1+z) \frac{d^2N_\gamma}{dzdE_p} P(f;E_p,z) \left[\frac{dN_{\Gamma_1}(E_0(1+z))}{dE}+\frac{dN_{\Gamma_2}(E_0(1+z))}{dE}\right] e^{-\tau(E_0,z)} dE_p dz\,,
\end{equation}
\end{widetext}
where $dN_\Gamma/dE$ is the spectrum of inverse Compton scattered radiation per electron of Lorentz factor, $\Gamma$, $P(f;E_p,z)$ is the probability that the pair production interaction at a given redshift, $z$, of a primary photon of energy $E_p$ will produce electron-type particles of energies $E_{e1} = f\times E_p$ and $E_{e2} = (1-f)\times E_p$, $\Gamma_1 = f\times E_p/mc^2$, $\Gamma_2 = (1-f)\times E_p/mc^2$, and $d^2N_\gamma/dzdE_p$ is the continuous spectrum of photons undergoing pair production interactions as a function of redshift and primary energy.

While numerical integration of Equation (\ref{eqn-cascadespec}) is, in principle, possible, doing so would be both complicated and computationally expensive\footnote{Furthermore, Equation (\ref{eqn-cascadespec}) is only valid for the case of EM cascades developed through pair production and inverse Compton interactions. While similar equations could possibly be derived for other types of interactions, one should take care that the derivation of equations of this nature relies on the accuracy of the description of continuous energy losses for the electrons.}.  Hence, it is desirable to make use of a Monte Carlo method, such as {\it Cascata}, that propagates photons from known spectra and redshift distributions and calculates the cascade spectrum of each photon, resulting in an \emph{observed} spectrum characterized by a discretized version of Equation (\ref{eqn-cascadespec}) (for derivation, see the Appendix),
\begin{equation}
\frac{dN_{\rm tot}}{dE_0} = \sum_{l=1}^{N_{\rm zbins}} (1+z_l) \Delta z_l \sum_{k=1}^{N_\gamma(z_l)} \sum_{j=1}^{N_e^k}\frac{dN_{\Gamma_j^k}}{dE}\,,
\end{equation}
where $N_{\rm zbins}$ is the number of redshift bins, $z_l$ is the redshift of the $l^{\rm th}$ bin, $\Delta z_l$ is the size of the $l^{\rm th}$ redshift bin, $N_{\gamma}(z_l)$ is the number of photons at $z_l$ undergoing pair production, $N_e^k = 2$ is the number or electrons produced by $k^{\rm th}$ photon undergoing pair production, and $dN_{\Gamma_j^k}/dE$ is the spectrum of inverse Compton photons produced by the $j^{\rm th}$ electron with Lorentz factor $\Gamma_j^k$.  We appropriately normalize the resulting spectrum in the final step.

In determining the cascade contribution to the EGRB from VHE photons from blazars, primary photon energies are randomly generated according to the \emph{absorbed} EGRB intensity (in observer's quantities), which is the differential amount of radiation that has been ``absorbed'' by the soft photon background: 
\begin{widetext}
\begin{equation}\label{eqn-abs}
I^a_E(E_0)
=\!\! \frac{c}{H_0} \frac{1}{4\pi E_f^2} \int_{-\infty}^{\infty} \!\!\! d\alpha\,p_L(\alpha) \!\! \left(\frac{E_0}{E_f}\right)^{-\alpha}\!\!\! \int_{z=0}^{z_{\rm max}} \! dz\,\frac{(1-e^{-\tau(E_0,z)})}{(1+z)^{\alpha}[\Omega_{\Lambda}+\Omega_m(1+z)^3]^{1/2}} \int_{L_{\gamma,{\rm min}}}^{L_{\gamma,{\rm max}}} \!\!\!\!\! dL_{\gamma}L_{\gamma}\rho_{\gamma}\,.
\end{equation}
\end{widetext}
This intensity is interpreted as the radiation that creates EM cascades.  As such, it will also be used to normalize the cascade spectra in the final step.

The \emph{absorbed} EGRB intensity is also used to generate the redshifts of the sources of the VHE photons by differentiating with respect to redshift: 
\begin{widetext}
\begin{equation}\label{eqn-absdz}
\frac{dI^a_E}{dz}(E_0) = \frac{c}{H_0}\frac{1-e^{-\tau(E_0,z)}}{4\pi E_f^2[\Omega_{\Lambda}+\Omega_m(1+z)^3]^{1/2}} \int_{-\infty}^{\infty} \! d\alpha\,p_L(\alpha)(1+z)^{-\alpha}\left(\frac{E_0}{E_f}\right)^{-\alpha} \!\!\! \int_{L_{\gamma,{\rm min}}}^{L_{\gamma,{\rm max}}} \!\!\!\!\! dL_{\gamma}L_{\gamma}\rho_{\gamma}\,.
\end{equation}
\end{widetext}
The created photons are then propagated until the redshift of interaction, randomly generated from the probability of interaction, $p = 1-\exp[-\tau(E_0,z)]$.

\subsection{The Cascade Formalism of {\it Cascata}}\label{subsec-propcode}

\begin{figure}[t]
\begin{center}
\resizebox{2.5in}{!}{\includegraphics[trim = 0 0 0 1mm, clip]{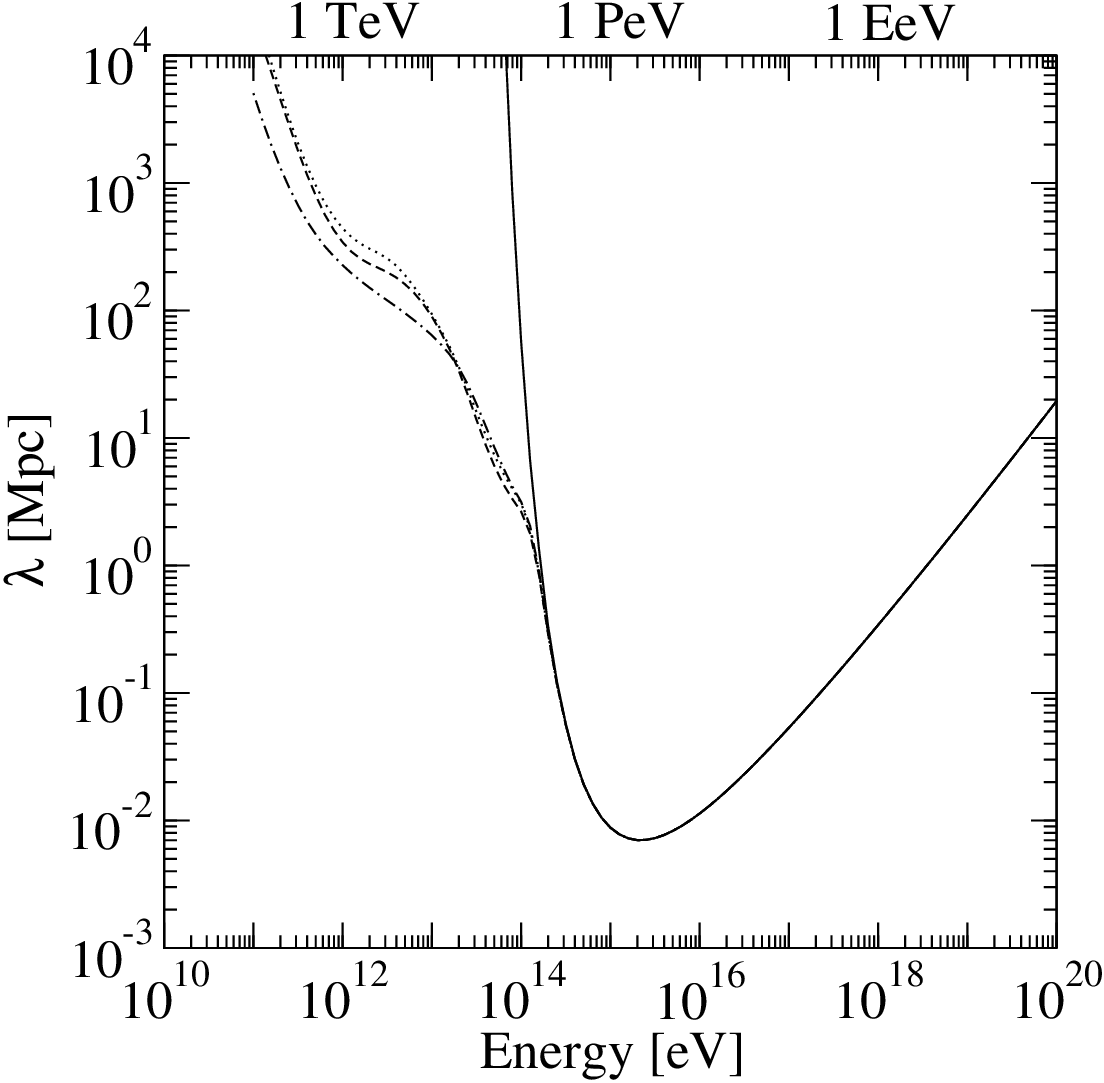}}
\vspace{0.1in}
\caption[Pair production mean free path for VHE photons.]{Pair production mean free path for VHE photons propagating through the present-day ($z = 0$) soft photon background as a function of energy for four different models of the background. \emph{Solid:} Soft photon background consisting of only the CMB. \emph{Dotted:} Soft photon background consisting of the CMB and the \citealt{gil09} model of the EBL. \emph{Dot-dashed:} Soft photon background consisting of the CMB and the \citealt{ste06} model of the EBL. \emph{Dashed:} For comparison, the soft photon background consisting of the CMB and the \citealt{kne07} model of the EBL.\label{fig-intrn}}
\end{center}
\end{figure}

In addition to generating primary photons, background photons with which the primary photons interact must also be generated according to the appropriate distributions \citep[see][for details]{pro86}.  For a given redshift, $z$, the background photon energies, $\epsilon$, may be sampled according to
\begin{equation}\label{eqn-probeps}
p(\epsilon) = \lambda_\gamma(E_p, z)n(\epsilon,z) \int_{-1}^{1} \!\!\! \frac{\sigma_{\gamma\gamma}(s)}{2}(1-\cos \theta)d(\cos \theta)\,,
\end{equation}
where $\lambda_\gamma(E_p, z)$ is the mean free path of a gamma-ray photon of energy, $E_p$, $n(\epsilon, z)d\epsilon$ is the proper photon number density (EBL plus cosmic microwave background (CMB)), $\theta$ is the angle between the paths of the gamma ray and the background photon, and $\sigma_{\gamma\gamma}(s)$ is the pair production cross section as a function of the square of the center of momentum (CM) energy, $s = E_{\rm CM}^2 = 2\epsilon E_p(1 - \cos \theta)$. The mean free path (see Figure \ref{fig-intrn})\footnote{The attenuation for photons of energies less than $10^{14}$ eV is dominated by the EBL. The various EBL models predict different amounts of UV radiation resulting in the differences seen in the figure in the mean free paths at energies of interest for IACTs and \emph{Fermi}.}, $\lambda_\gamma(E_p, z)$, of the gamma-ray photon is given by
\begin{equation}
\lambda^{-1}_\gamma(E_p, z) = \int^1_{-1} \frac{\left(1-\cos \theta \right)}{2} \int^{\infty}_{\epsilon_{\rm th}} n(\epsilon, z) \sigma_{\gamma\gamma}(s) d\epsilon d(\cos \theta)\,.
\end{equation}
where $\epsilon_{\rm th} = 2m^2_ec^4/E_p(1-\cos \theta)$. The pair production cross-section is given by
\begin{equation}
\sigma_{\gamma\gamma}(s) = \frac{3}{16}\sigma_{T}(1-\beta^2)\left[(3-\beta^4)\ln \frac{1+\beta}{1-\beta} - 2\beta(2-\beta^2) \right]\,,
\end{equation}
where $\beta c = (1-4m^2_ec^4/s)^{1/2}c$ is the magnitude of the velocity of the electrons in the CM frame. The angle between the paths of the two photons can be determined by sampling the square of the CM energy, $s$, from Equation (\ref{eqn-probeps}) differentiated with respect to $s$ \citep{pro86}\footnote{In differentiating Equation (\ref{eqn-probeps}) with respect to $s$, we are, in fact, differentiating only with respect to $\cos \theta$, where again $s = 2\epsilon E_p (1-\cos \theta)$.}
\begin{equation}
p(s) \propto s\sigma_{\gamma\gamma}(s)\,.
\end{equation}
The scattering angle, $\alpha$, of the secondary electrons in the CM frame is determined from the differential pair production cross section:
\begin{equation}
p(C) = \frac{d\sigma_{\gamma\gamma}}{d\cos \alpha} \propto \left[\frac{1-\beta^4C^4+2\beta^2(1-\beta^2)(1-C^2)}{(1-\beta^2C^2)^2}\right]\,,
\end{equation}
where $C = \cos \alpha$. The energies of the electrons in the lab frame are then completely specified by the energy of the propagating gamma ray, the randomly generated parameters of the pair production interaction ($\epsilon$, $\theta$, and $\alpha$), and relativistic kinematics\footnote{The procedure for determining the energies of the two secondary electrons created through the interaction between the two primary photons involves boosting from the lab frame to the CM frame, determining the energies and momenta of the two secondary electrons in the CM frame, and then boosting back to the lab frame.}: 
\begin{equation}
E_{e1} = \gamma c\left[\frac{E_{\rm CM}}{2} + \beta |\vec{p}\,'_{\! e1}|\cos(\alpha+\psi)\right]
\end{equation}
and
\begin{equation}
E_{e2} = \gamma c\left[\frac{E_{\rm CM}}{2} - \beta |\vec{p}\,'_{\! e2}|\cos(\alpha+\psi)\right]\,,
\end{equation}
where $\gamma$ and $\beta$ determine the boost from the lab frame to the CM frame, $\psi$ is the angle of the photons with respect to the $x$-axis in the CM frame, $|\vec{p}\,'_{\! e1}|=(E^2_{\rm CM}/4c^2 - m_e^2c^2)^{1/2}$ and $|\vec{p}\,'_{\! e2}|=(E^2_{\rm CM}/4c^2 - m_e^2c^2)^{1/2}$ are the magnitudes of the CM momenta of the secondary particles.

Finally, the electrons scatter background photons through the inverse Compton process resulting in a scattered photon spectrum given by (see the Appendix)
\begin{equation}
\frac{dN_{\Gamma}}{dE} = \int \!\! \frac{d^3N_{\Gamma'}}{dtdEd\epsilon}\left|\frac{dt}{dE_e}\right|dE_ed\epsilon\,,
\end{equation}
where $d^3N_{\Gamma'}/dtdEd\epsilon$ is differential scattered photon spectrum given by Equation (\ref{eqn-diffscatspec}) \citep{blu70} and $dE_e/dt$ is the electron loss rate\footnote{Note that this process must be repeated iteratively if VHE electrons are produced.}.

\section{Inputs}
\label{cha:inputs}

\subsection{Blazar Gamma-ray Luminosity Functions}\label{subsec-blglf}

Determining the blazar GLF from observations relies on the ability to associate gamma-ray blazars with lower energy counterparts for which redshifts can be measured.  However, identifying gamma-ray-loud blazars can be complicated due to the large positional error circles of GeV gamma-ray experiments.  Blazar variability and low number statistics further complicate the determination of the GLF.  Thus, while the blazar GLF is probably one of the most studied and debated properties of the gamma-ray-loud blazar population, it remains, to this day, uncertain.

In gamma-ray observations, blazar GLFs are constructed from luminosity functions based on survey data taken at wavelengths for which at least some of the observational uncertainties are not as challenging. The most uncertain parameters (e.g., normalization due to relativistic beaming and the faint-end slope) are then fitted to gamma-ray data \citep[see e.g.,][NT06]{ss96a,gio06}.  Such an approach raises a number of concerns: 
\begin{enumerate}
\item[(1)] Not all blazars are necessarily explained by the same emission process, and different types of blazars (i.e., BL Lac-like objects and FSRQs) could form separate populations with respect to emission \citep{sik02,bot07} and, hence, require separate luminosity functions \citep{muc00,der07}.
\item[(2)] Flaring blazars (and different \emph{types} of flaring blazars) could also form separate populations with respect to emission and require separate luminosity functions \citep{ss96a}.
\item[(3)] Many unidentified gamma-ray sources could be \emph{unidentified blazars}, which, without identification, would result in underestimating the normalization of gamma-ray blazars with respect to low-energy blazars and the introduction of uncertainties in the redshift distribution of resolved blazars.
\end{enumerate}
Alternative approaches have been proposed \citep[see, e.g.,][]{der07}, but they are also subject to uncertainty.

With the availability of \emph{Fermi} data, many more blazars will be observed, and at least some of the aforementioned uncertainty will be alleviated.  In fact, with the release of the first year data approaching, predictions for the \emph{Fermi} source counts will be directly tested.  Furthermore, if the collective intensity of blazars does, in fact, constitute a substantial contribution to the EGRB, then \emph{Fermi} will observe an absorption feature at the highest energies \citep{ven09}.  The shape of such an absorption feature is highly sensitive to the blazar GLF and, hence, can constrain the evolution of blazars\footnote{It should be noted that the shape of the absorption feature is also highly sensitive to the EBL (see Section \ref{subsec-EBL}) through which VHE gamma rays propagate.  As such, the absorption feature can also constrain EBL models.}.  While \emph{Fermi} data from the first few months of observations of AGNs have been released \citep{lat09a}, the sources listed in the catalog represent high sensitivity detections ($\sim 10\sigma$) and do not constitute a complete sample. Thus, for the purposes of this paper, we make use of the best-fit pure luminosity evolution (PLE) and luminosity-dependent density evolution (LDDE) models of the likelihood analysis of NT06. Calculations based on luminosity functions, based on \emph{Fermi} first year data, will be addressed in a future publication.

\subsection{Blazar Spectra}\label{subsec-blsp}

As with the GLF, determining the contribution of the collective intensity of any population of gamma-ray emitters to the EGRB requires precise knowledge of their spectra.  If all blazar spectra were described by the same power law, then the unabsorbed collective intensity would also follow that same power law.  If, on the other hand, the spectral indices of the population of blazars form a distribution with a given finite spread (the SID), the spectrum of the unabsorbed collective intensity will exhibit curvature due to the increasing relative importance at high energies of blazars with harder spectral indices \citep{ss96a,pv08}.  Since the radiation at the energies which are likely to initiate EM cascades is most sensitive to the effect of the spread in the SID, it is essential that the blazar SID be determined.

In the analysis of blazar spectral indices, we make the critical assumption that the spectral index of a blazar is independent of its luminosity \citep[as indicated in][]{vp07}. Under this assumption, the incompleteness of the \emph{Fermi} LBAS with respect to luminosity does not hinder the determination of the SID even though it complicates the extraction of blazar GLFs. Thus, for the purposes of this paper, we determine the blazar SIDs from the \emph{Fermi} LBAS blazars following the likelihood procedure outlined in \citet{vp07}.  For the \emph{Fermi} LBAS blazars, we find the parameters of the maximum likelihood (Gaussian) SID to be ($\alpha_0 = 2.23$, $\sigma_0 = 0.26$).  For the BL Lacs, we find ($\alpha_0 = 1.97$, $\sigma_0 = 0.175$); and for the FSRQs, we find ($\alpha_0 = 2.37$, $\sigma_0 = 0.14$)\footnote{Note that the best-fit spreads are slightly smaller than that indicated by \citet{lat09a}.  This is because \citet{lat09a} calculate the rms of the \emph{observed} blazar spectral indices, which does not account for measurement uncertainty in each individual spectral index.}.
We also correct for the sample bias introduced in using a flux-limited catalog\footnote{Blazars observed in a flux-limited survey will be biased toward high-luminosity, low-redshift blazars.  Low-luminosity, high-redshift blazars that are most likely to appear in a gamma-ray survey are those with spectral indices that are hard with respect to the population as a whole.  However, in this case, the corrections will only be approximate since we are not calculating them from luminosity functions based on LBAS blazars. Furthermore, the LBAS is not exactly a flux-limited survey due to the variation in the diffuse intensity with galactic latitude \citep{lat09b}.} \citep[for discussion, see][]{ven09}.


Finally, while it may be true that blazar spectra appear as single power laws in the GeV band, it may not be so beyond the GeV band (or, at least, it may not be the same power law).  Observations by IACTs demonstrate that blazar spectra are steeper at VHEs \citep{alb08}, and even the \emph{Fermi} LBAS provides a few examples of blazar spectra that appear to break at energies greater than $\sim 10$ GeV.  In the calculation of the blazar contribution to the EGRB including cascades, one would ideally include the effects of possible breaks in blazar spectra.  However, doing so is quite complicated as it is not clear whether the breaks are due to changes in the spectra endemic to the sources or, at least in part, to attenuation, particularly for high-redshift blazars. Even if the breaks do reflect intrinsic changes in the source spectra, the cause(s) of the breaks is not, as yet, well understood, and hence the intrinsic spectral breaks and the energies of these breaks are uncertain.  Additionally, comparisons between blazar GeV spectral indices and TeV spectral indices require simultaneous observations of blazars by \emph{Fermi} and IACTs, and as yet, the sample size for such observations is quite small\footnote{However, there is considerable interest within the gamma-ray community to perform such coordinated observations, and even though the different natures of the involved experiments ensure that they will be performed on a case-by-case basis, the sample size of multi-wavelength observations will grow with time.}. Finally, the observations provided by the IACTs and the spectral breaks demonstrated in the LBAS do not constitute a flux-limited catalog, and as such, the implications of these observations for the blazar population as a whole remain unclear. In considering spectral indices from a flux-limited catalog, correcting for the observational bias introduced by the overrepresentation of harder spectral indices in the SID is relatively straightforward \citep{ven09}, but this is not the case with a catalog that is not flux-limited. Thus, in light of the uncertainty surrounding blazar spectral breaks, we consider the study of their effects to be out of the scope of this paper; hence, for the purposes of this analysis, we assume that blazar spectra remain unbroken at VHEs. As such, the calculated cascades will be larger than expected for blazar spectra which soften at VHEs. 

We will return to the question of blazar spectral breaks in an upcoming publication. In the meantime, we note that \emph{Fermi} measurements of the EGRB and theoretical determinations of the blazar contribution to the EGRB that include cascade radiation can provide constraints on blazar spectra at VHEs within the context of the blazar population as a whole, which would be, in effect, \emph{complimentary} to the constraints provided by IACTs on individual blazars.

\subsection{Extragalactic Background Light}\label{subsec-EBL}

The spectral energy distribution of the cosmic background of infrared, optical, and ultraviolet light (the EBL) encodes much information about the astrophysical and cosmological processes in the universe. The EBL intensity originates from starlight (at optical, ultraviolet, and near-infrared wavelengths) and reradiated thermal dust emission (far-infrared) in galaxies.  Hence, the EBL is directly connected to the evolution with cosmic time of structure in the universe.  However, the nature of key contributors to the EBL is largely uncertain (e.g., the evolution of the cosmic star formation rate; the amount of UV radiation from young, massive stars that escapes from the surrounding gas; the amount of re-emission by dust in galaxies).  Furthermore, observations are hindered by emission from our own galaxy \citep[for detailed review, see][]{hau01}.  Nevertheless, despite the degree of complication, there are several approaches to overcoming the observational and theoretical uncertainties and, ultimately, modeling the EBL \citep[for more detailed discussion, see][]{kne02}.
\begin{enumerate}
\item[(1)] Forward evolution of structure formation and the resulting luminosity functions \citep[e.g.,][]{pri08,gil09};
\item[(2)] Backward evolution of observed local luminosity functions to higher redshifts, most notably at infrared wavelengths \citep[e.g.,][]{ste06,kne02,kne04,kne07};
\item[(3)] Evolution determined from direct observations over some range in redshift such as the global star formation rate inferred from deep galaxy surveys or tracers of cosmic chemical evolution convolved with galaxy spectral energy distributions determined from population synthesis models \citep[e.g.,][]{ste06,kne02,kne04,kne07}.
\end{enumerate}

While progress is being made on all fronts, each method remains, as yet, subject to its own uncertainties, so it is not obvious which is the best one to use in propagation studies.  Thus, we adopt two models that are representative of the current state of the field and bracket the possibilities: the \citet{ste06} EBL model and the \citet{gil09} EBL model \citep[for more information, see][and references therein]{ven09}. The most notable distinction between these two models is the amount of UV radiation predicted.  Since the Stecker model does not include the effects of dust extinction, it predicts a higher amount of UV radiation than the Gilmore model. Other prevalent models in the literature, such the \citet{kne07} \emph{high-UV} EBL model, the \citet{kne07} \emph{best-fit} EBL model, and the \citet{fra08} EBL model, yield results similar to those of the Stecker and Gilmore models (at least in the analysis presented in this paper).

In propagating VHE gamma rays, we make extensive use of the optical depth of VHE photons due to pair production interactions computed from the proper photon number density, $n(\epsilon,z)d\epsilon$:
\begin{equation}\label{eqn-optical_depth}
\tau\left(E_0,z\right)  =  \int_{0}^{z}\!\!\! dz'\frac{dl}{dz'} \!\! \int_{0}^{2\pi}\!\!\! \sin\theta d\theta \!\! \int_{\epsilon_{th}}^{\infty} \!\!\!\! d\epsilon\,\, n\left(\epsilon,z'\right)\,\sigma(s)\,,
\end{equation}
where
\begin{widetext}
\begin{equation}
\int_{0}^{z}dz'\frac{dl}{dz'} =  \int_{0}^{z} \!\!\! dz'\,\frac{c}{H_{0}\left(1+z'\right)}\left[\left(1+z'\right)^{2}\left(1+\Omega_{M}z'\right)-z'\left(2+z'\right)\Omega_{\Lambda}\right]^{-1/2}
\end{equation}
\end{widetext}
is the look-back time.

\section{Results}
\label{cha:results}

In this section, we discuss the effects of EM cascade radiation from the propagation of VHE photons from blazars through the EBL on the overall collective blazar intensity and the implications for their contribution to the EGRB. In each plot, we show the overall collective blazar intensity, the contribution to the overall collective intensity from the collective intensity of \emph{intrinsic} blazar spectra, and the contribution to the overall collective intensity from cascade radiation arising from propagation. We also discuss the dependence of the cascade radiation on the blazar GLF, the blazar SID, and the model of the EBL.

\begin{figure}[t]
\begin{center}
\resizebox{2.5in}{!}
{\includegraphics[trim = 0 0 0 1mm, clip]{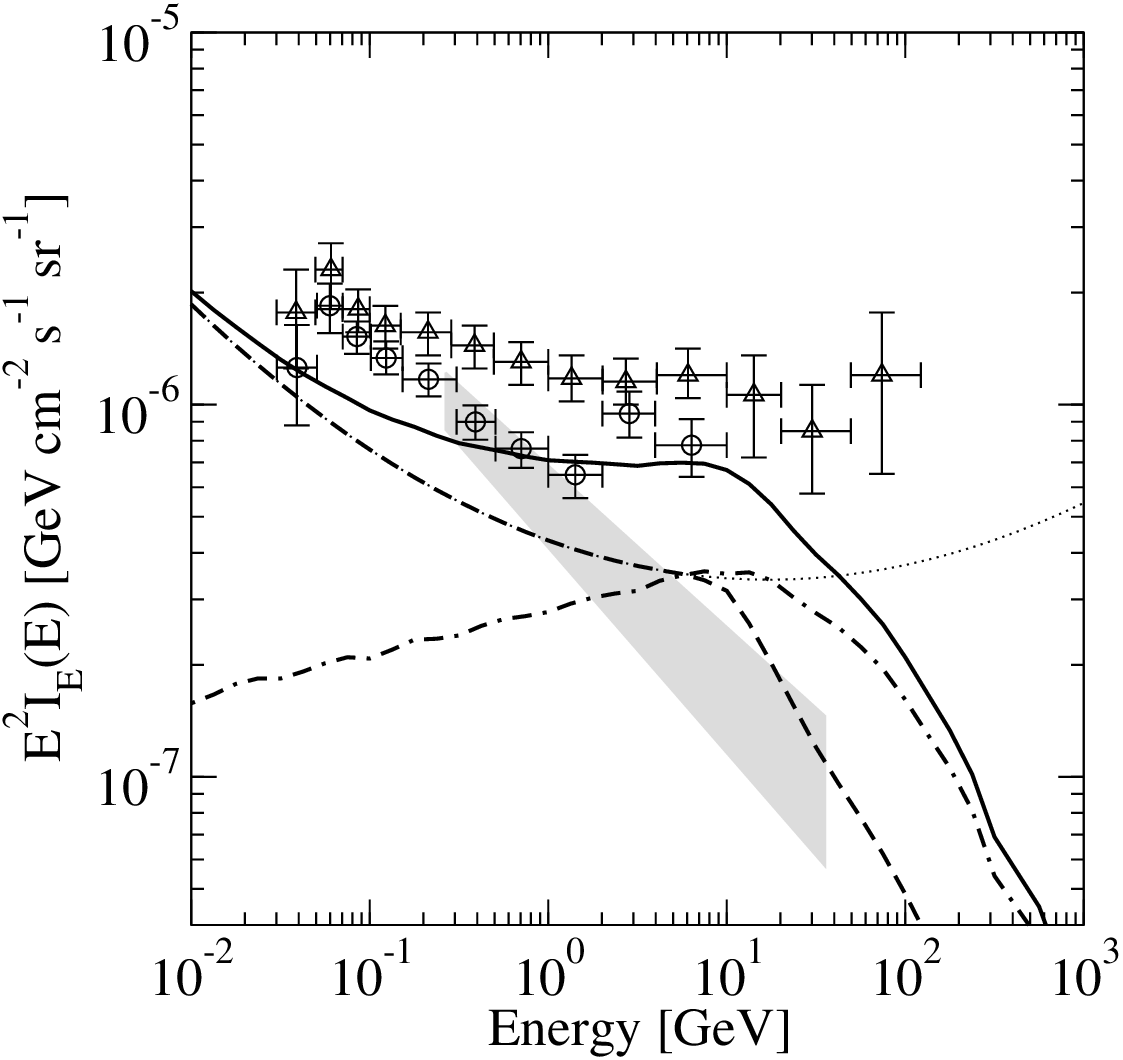}}
\resizebox{2.5in}{!}
{\includegraphics[trim = 0 0 0 1mm, clip]{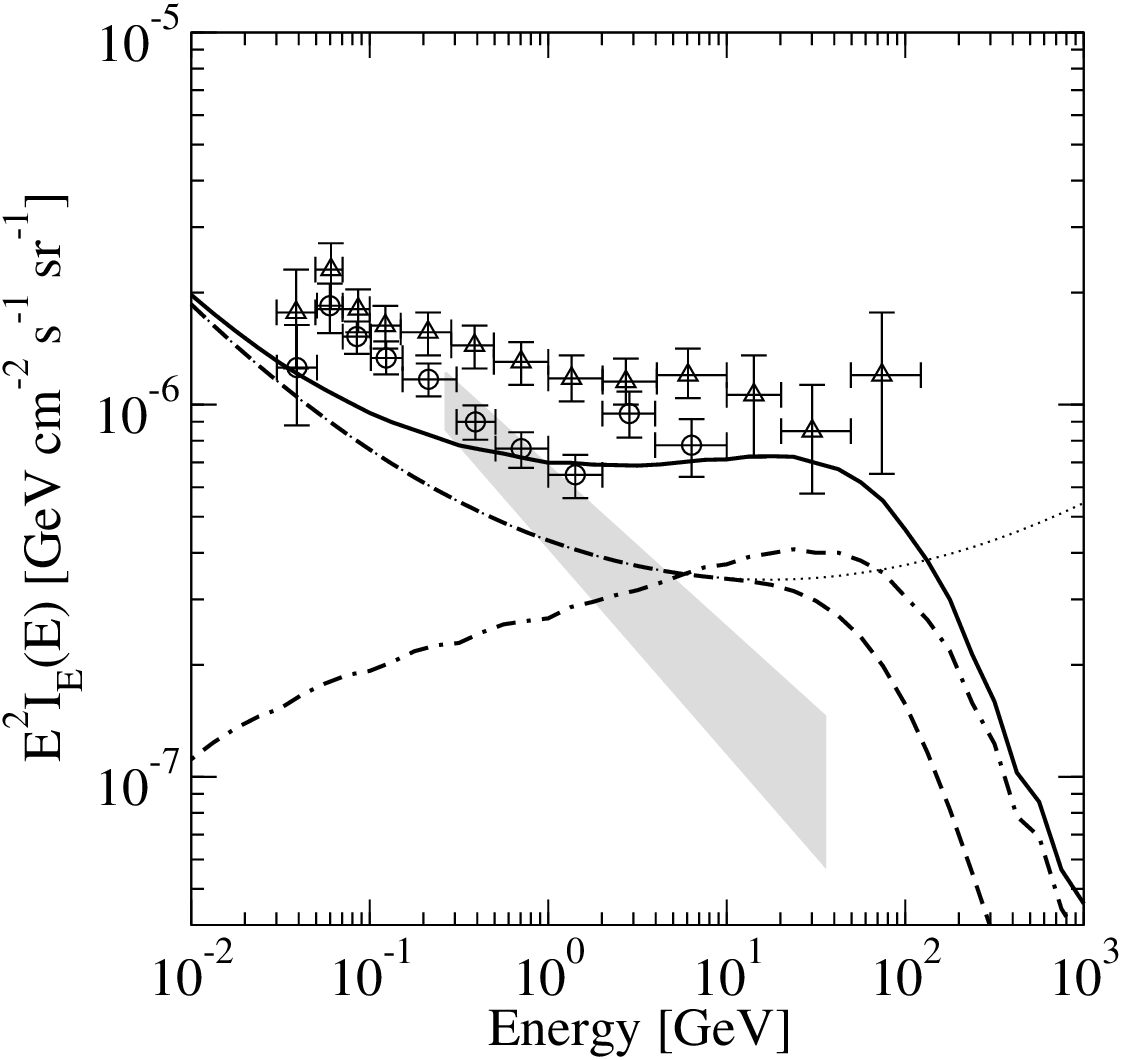}}
\caption{The blazar contribution to the EGRB including EM cascade radiation from the propagation of VHE photons through the EBL assuming the PLE model of the blazar GLF.  \emph{Top:}  The calculation assuming the Stecker EBL model. \emph{Bottom:} The calculation assuming the Gilmore EBL model. \emph{Solid lines:} The total blazar contribution, including EM cascade radiation. \emph{Dashed lines:} The collective intensity of \emph{intrinsic} blazar spectra without cascades but including absorption. \emph{Dot-dashed lines:} The intensity from EM cascades, including absorption. \emph{Dotted lines:} The collective intensity not including absorption and cascade radiation. \emph{Square data points:} \citet{sre98} EGRET EGRB. \emph{Circular data points:} \citet{str04} EGRET EGRB. \emph{Grey band:} Preliminary \emph{Fermi} EGRB with statistical and instrumental systematic uncertainties.\label{fig-BlazarEGRBPLE}}
\end{center}
\end{figure}

\begin{figure}[t]
\begin{center}
\resizebox{2.5in}{!}
{\includegraphics[trim = 0 0 0 1mm, clip]{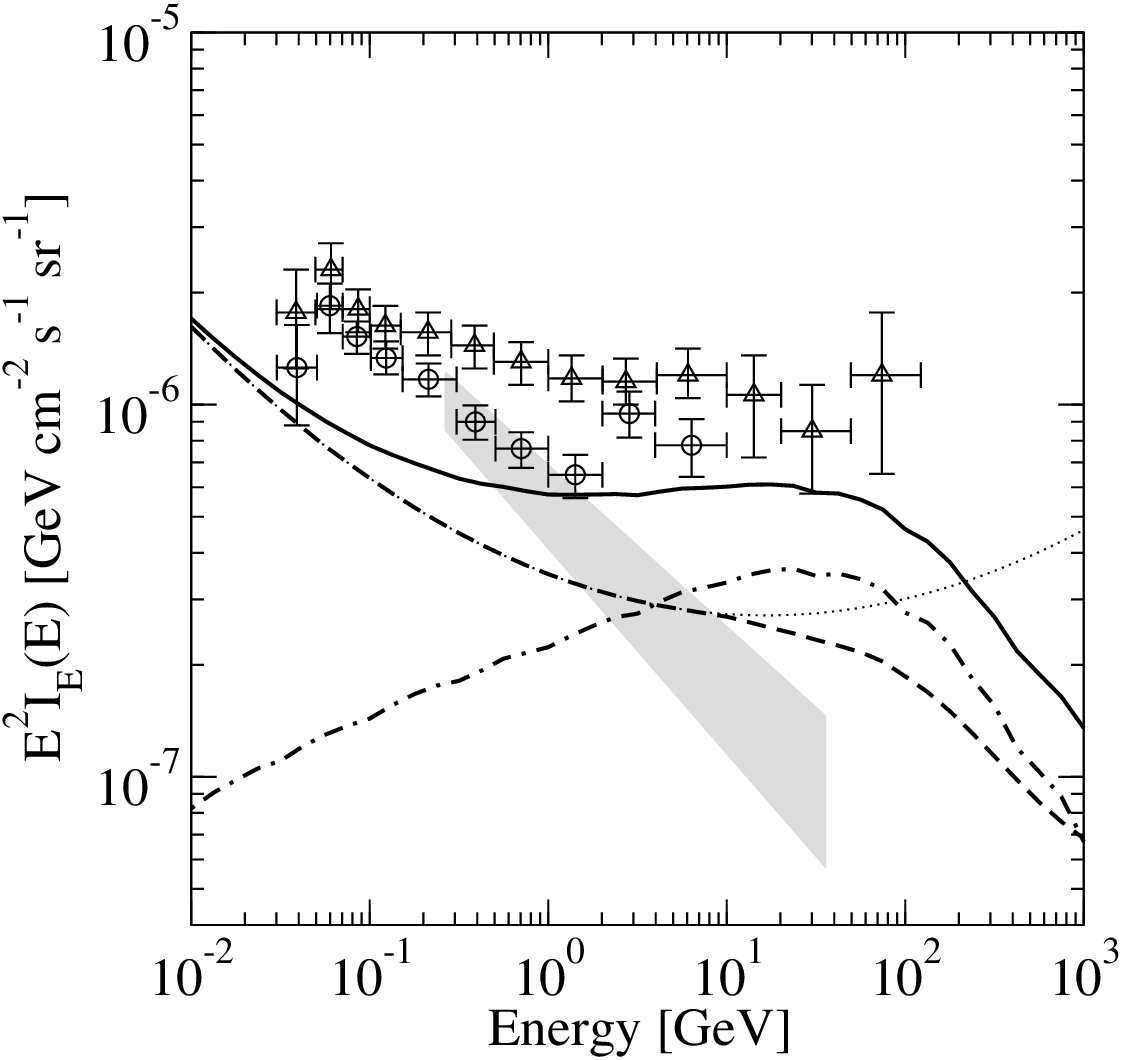}}
\resizebox{2.5in}{!}
{\includegraphics[trim = 0 0 0 1mm, clip]{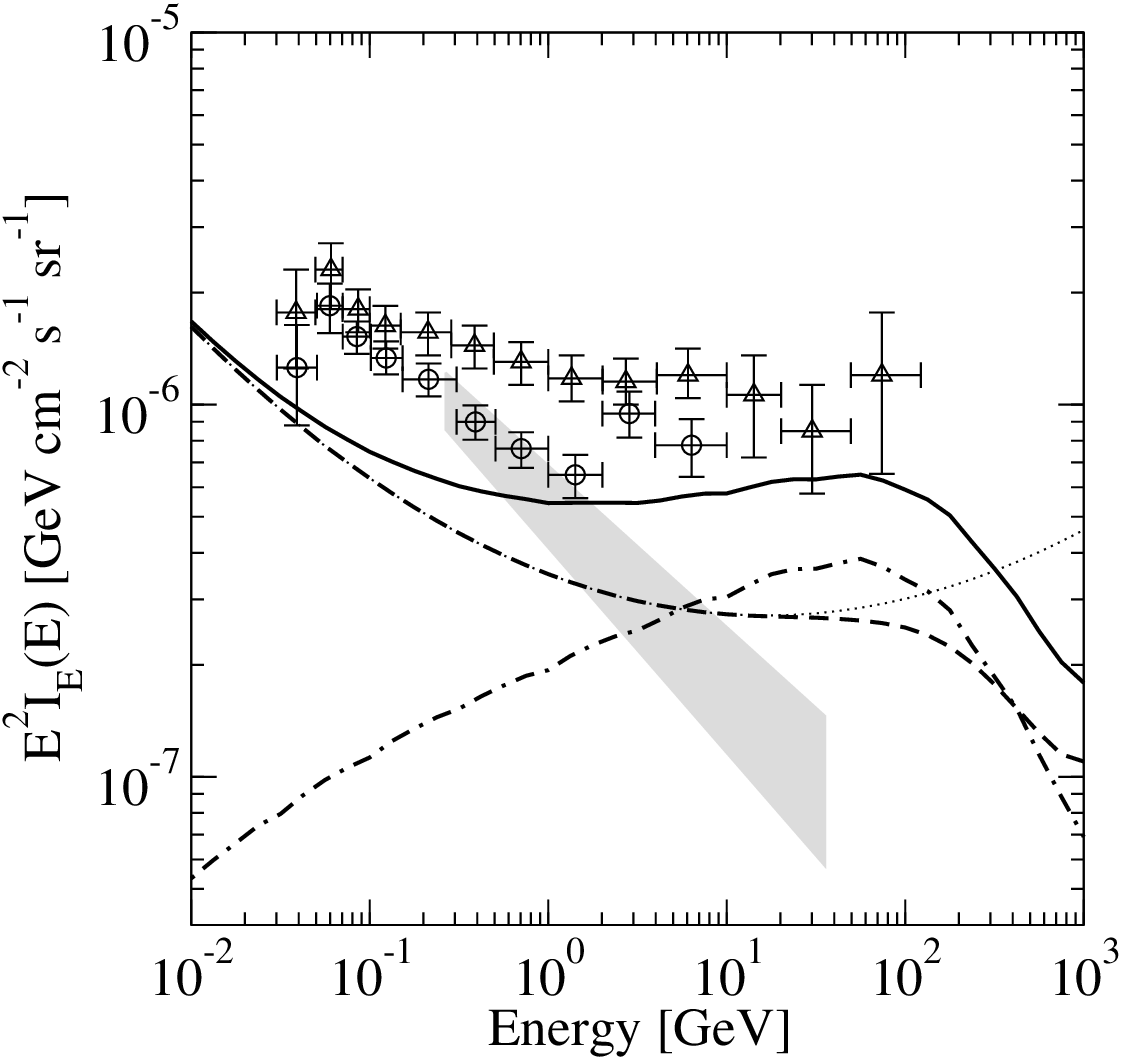}}
\caption{Same as Figure \ref{fig-BlazarEGRBPLE} except for the LDDE model of the blazar GLF.\label{fig-BlazarEGRBLDDE}}
\end{center}
\end{figure}

The blazar contributions to the EGRB including cascade radiation for two models of the blazar GLF are plotted in Figures \ref{fig-BlazarEGRBPLE} and \ref{fig-BlazarEGRBLDDE}. For both GLFs, the calculation was performed for two separate models of the EBL (solid lines represent the total intensity from blazars, including cascade radiation, for each of the four possibilities). Figure \ref{fig-BlazarEGRBPLE} presents the calculation assuming the NT06 PLE model of the GLF, and Figure \ref{fig-BlazarEGRBLDDE} presents the calculation assuming the NT06 LDDE model of the GLF\footnote{Note that for the best-fit parameters of these luminosity functions the collective intensity of blazars (not including absorption and cascade radiation) comprises $\sim 50$\% of the EGRB.  However, in several of the cases presented in NT06, parameters for which unresolved blazars can account for $100$\% of the background are within the $1\sigma$ contours. It should further be noted that with the parameters provided by NT06, the LDDE model predicts an overall lower intensity than the PLE model.}.  The top panels present the calculations assuming the \citet{ste06} EBL model, while the bottom panels present the calculations assuming the \citet{gil09} EBL model. All calculations are made assuming the parameters of the SID calculated based on the \emph{Fermi} LBAS (see Section \ref{subsec-blsp})\footnote{There is some uncertainty in the determination of the parameters of the blazar spectral index, which will result in uncertainty in the overall shape of the spectrum of the unresolved blazar intensity. The already greatly improved blazar number statistics in the \emph{Fermi} LBAS has reduced this uncertainty; hence, for the purposes of this paper, we simply analyze the best-guess spectra.}. For comparison, we have also plotted the \citet{sre98} determination of the EGRB as measured by the EGRET (square data points with statistical error bars), the \citet{str04} determination of the EGRET EGRB (circular data points with statistical error bars), a gray band representing the \emph{Fermi} preliminary EGRB (including statistical and instrumental systematic uncertainties) from the first 9 months of data\footnote{As presented by M. Ackermann at TeV Particle Astrophysics 2009}, the collective intensity of \emph{intrinsic} blazar spectra assuming no absorption (dotted lines), the collective intensity of \emph{intrinsic} blazar spectra including only absorption of the highest energy photons (dashed lines), and the cascade radiation (dot-dashed lines).  In this analysis, we include intrinsic emission up to $100$ TeV.

As evident, the amount of cascade radiation created in the propagation of VHE photons can be substantial, depending on the shape of the collective intensity of intrinsic blazar spectra (for examples of cascade radiation resulting from other shapes, see below).  For the shape determined by the SID indicated by \emph{Fermi} data, the cascade radiation significantly alters both the amount and the spectral shape of the overall intensity. In fact, for model GLFs that predict the contribution of intrinsic blazar spectra to comprise only $\sim 50$\% of the EGRB, the addition of cascade radiation greatly increases the overall contribution from blazars, at least at the highest energies. At lower energies, the cascade radiation drops off slowly over several decades in energy, and the collective intensity of intrinsic blazar spectra rises quickly; thus, the contribution of the cascade radiation to the overall collective blazar intensity is much less at lower energies.

Also seen in the figure is the contrast in appearance between the PLE model of blazars and the LDDE model, most notably at the highest energies.  As noted in \citet{ven09}, the collective blazar intensity is much more distributed with respect to redshift in the PLE model than in the LDDE model, in which the radiation tends to be concentrated at lower redshifts (see their Figure 2).  Hence, just as the absorption feature in the collective intensity of intrinsic blazar spectra is much more prominent in the PLE model, so is it in the cascade radiation calculated from this model.  Thus, not surprisingly, while there is more cascade radiation below $10$ GeV in the PLE model than in the LDDE model, above $10$ GeV, the cascade radiation in the LDDE model overtakes that of the PLE model.  Since there is more high-energy absorption of the collective intensity from intrinsic blazar spectra in the PLE model, the cascade radiation in the PLE model dominates the intensity at energies beyond the EGRET energy range, contributing nearly 4 times as much radiation as the contribution from intrinsic blazar spectra at $100$ GeV.  Not so for the LDDE model, in which the cascade radiation contributes at most $50$\% more than that of the intrinsic blazar spectra at $100$ GeV.

In addition to considering various blazar GLF models, we also investigate the impact of the EBL model.  In both models, the effect of the cascade radiation is to flatten the overall collective intensity from blazars.  However, at the highest energies, there is less absorption in the Gilmore model than in the Stecker model. Below $10$ GeV, the cascade radiation for the two models is roughly comparable, though there is more radiation at the lowest energies for the Stecker model.  Above $10$ GeV in the PLE model of blazars, there is more cascade radiation for the Gilmore model than for the Stecker model due to the enhanced absorption in the Stecker model.  As a result, the spectra at high energies are vastly different.  As noted in \citet{ven09}, such an absorption feature is much more prominent in the Stecker case than in the Gilmore case owing to the higher UV background in the Stecker EBL model.  Since the pair production cross section as a function of the CM energy peaks at twice the electron mass, one would expect that gamma-ray photons of energies $\sim$ tens of GeV are most likely to interact with UV background photons. Thus, unsurprisingly, models with high UV backgrounds will result in more suppression at high energies. On the other hand, in the LDDE model, while there is still more cascade radiation at high energies for the Gilmore model than for the Stecker model, the resulting distinction between the two models is less prominent than in the PLE model owing to the preferential distribution of relatively high-flux blazars at low redshifts.

\begin{figure}[t]
\begin{center}
\resizebox{2.5in}{!}
{\includegraphics[trim = 0 0 0 1mm, clip]{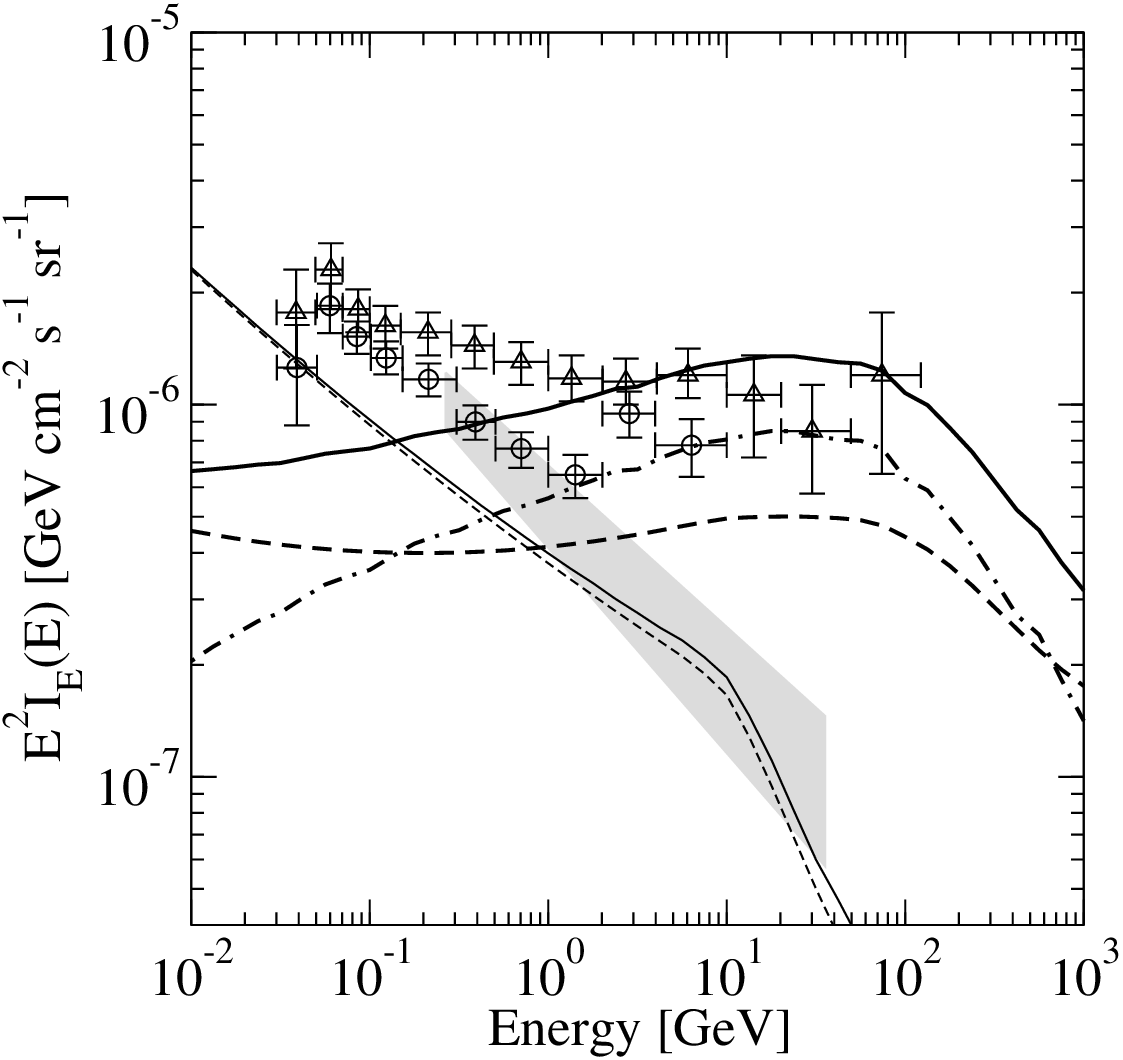}}
\resizebox{2.5in}{!}
{\includegraphics[trim = 0 0 0 1mm, clip]{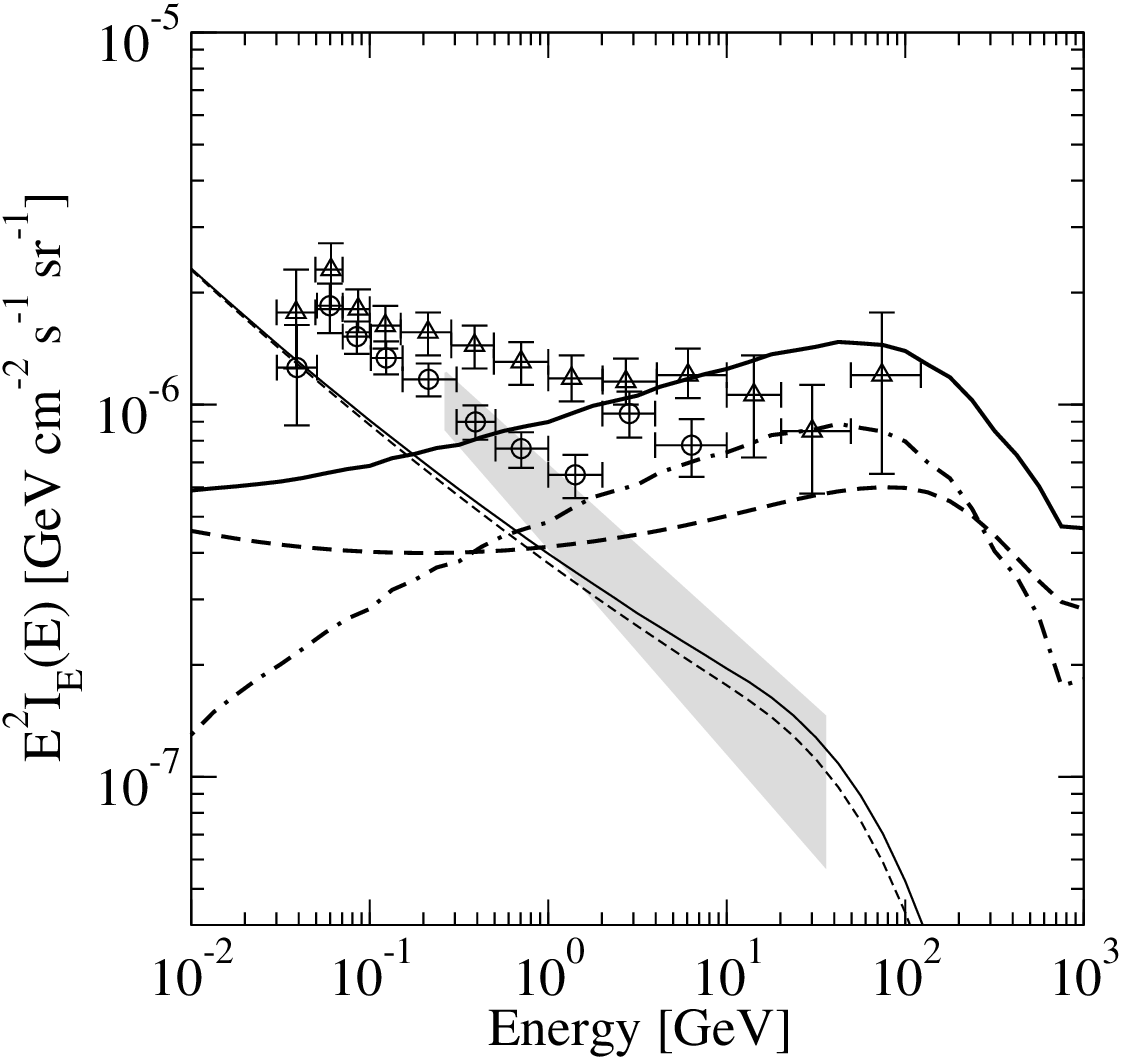}}
\caption{The blazar contribution to the EGRB including EM cascade radiation assuming separate emission properties for BL Lacs and FSRQs. \emph{Thick lines:} BL Lac contribution assuming the LDDE model and their \emph{Fermi}-determined SID, ($1.97$, $0.175$). \emph{Thin lines:} FSRQ contribution assuming the PLE model and their \emph{Fermi}-determined SID, ($2.37$, $0.14$). \emph{Data points, grey band, solid lines, dashed lines, and dot-dashed lines:} Same as in Figure \ref{fig-BlazarEGRBPLE}.\label{fig-FSRQBLL}}
\end{center}
\end{figure}

We also investigate the possibility that BL Lacs and FSRQs form separate populations with respect to emission and evolution by propagating photons from distinct GLFs and SIDs for the two populations (Figure \ref{fig-FSRQBLL}). As both the third EGRET catalog and the LBAS indicate that the BL Lacs tend to be more concentrated at low redshifts as compared to the FSRQs, we assume their evolution to follow the LDDE model.  By contrast, we assume the FSRQs to follow the PLE model\footnote{In principle, in separating the blazar population into its spectral sub-populations, the parameters for the GLF models would have to be refitted to the measured distributions of their corresponding sub-populations. However, for the sake of discussion, we retain the parameters for each GLF model as determined by NT06.}. For the SIDs, we assume the parameters determined from the \emph{Fermi} LBAS for the respective sub-populations (see Section \ref{subsec-blsp}).

\begin{figure}[t]
\begin{center}
\resizebox{2.5in}{!}
{\includegraphics[trim = 0 0 0 1mm, clip]{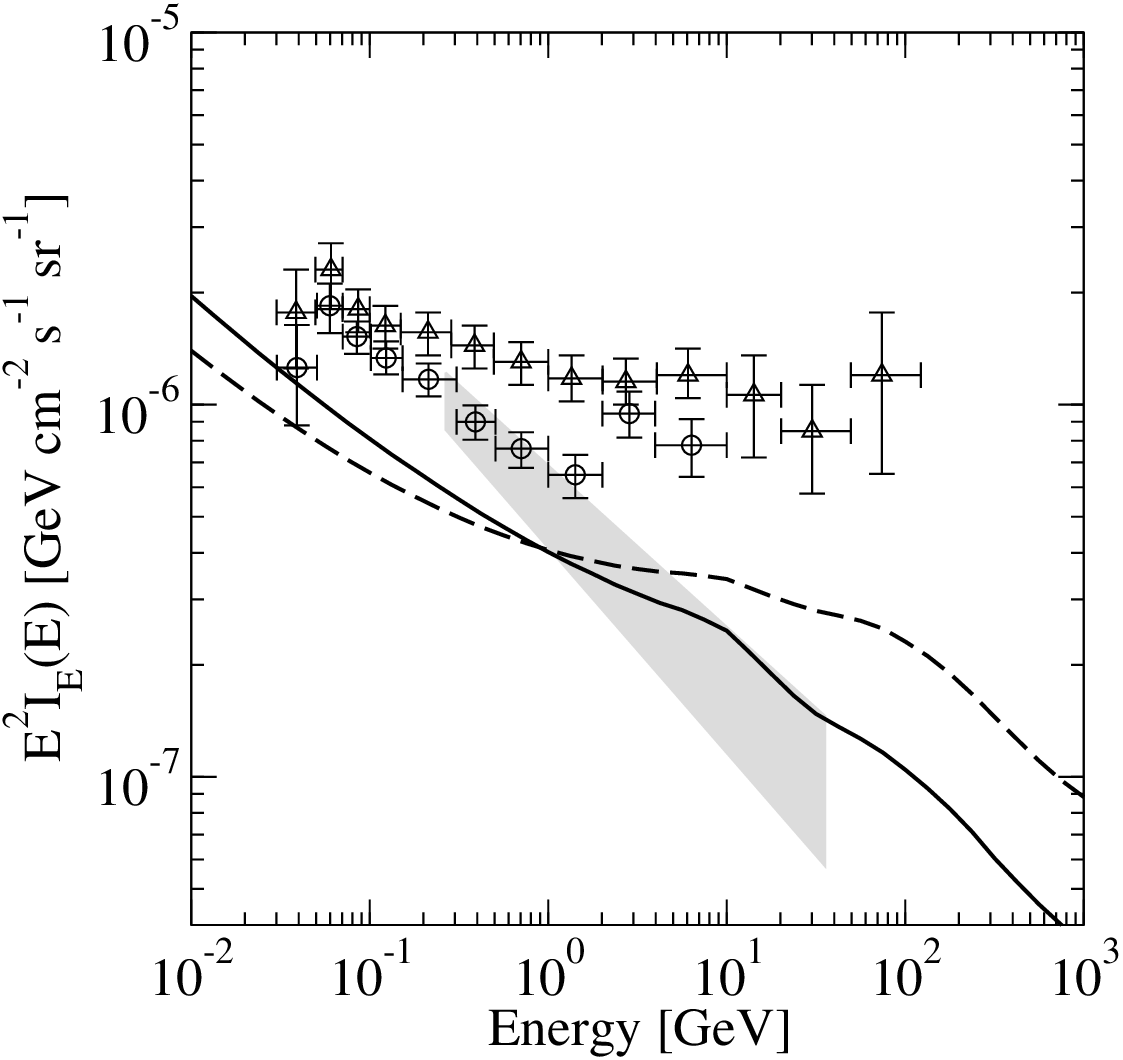}}
\resizebox{2.5in}{!}
{\includegraphics[trim = 0 0 0 1mm, clip]{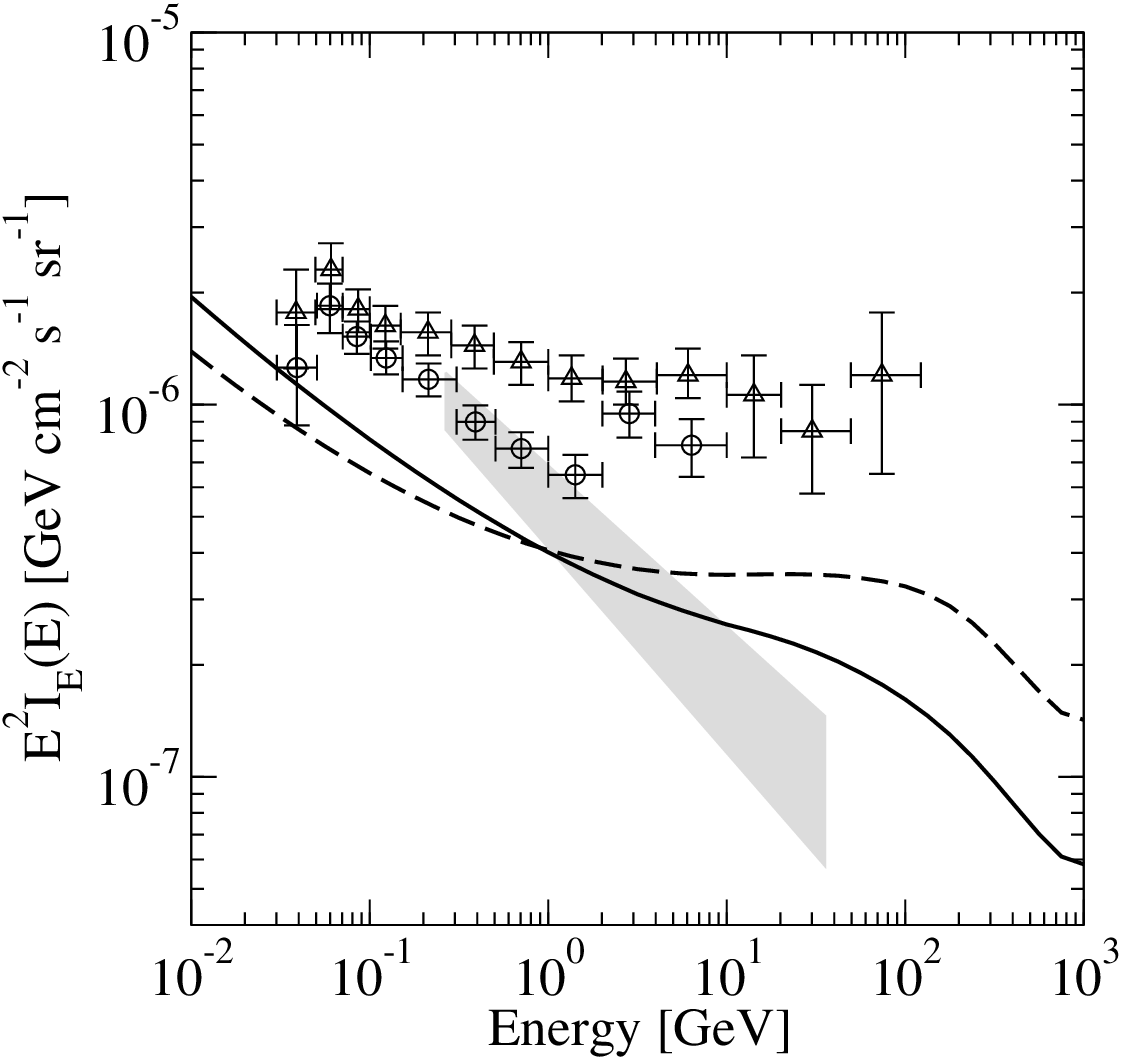}}
\caption{The superposition of the contributions to the EGRB of FSRQs and BL Lacs including EM cascade radiation assuming separate GLFs and SIDs for the subpopulations. \emph{Solid:} The contribution assuming FSRQs comprise $80$\% of the blazar population and BL Lacs comprise $20$\%. \emph{Dashed:} The contribution assuming equal numbers of FSRQs and BL Lacs. \emph{Data points and grey band:} Same as in Figure \ref{fig-BlazarEGRBPLE}.\label{fig-FSRQBLLadded}}
\end{center}
\end{figure}

The hard spectra of BL Lacs cause the resulting cascade radiation to be considerable relative to that of the collective intensity from intrinsic spectra even though their collective radiation tends to be concentrated at low redshifts in this model. In fact, the total intensity from BL Lacs could dominate the EGRB at high energies and even overproduce it, though it does not explain the lower energy intensity. By contrast, the relatively steep spectra of FSRQs are not conducive to much cascade radiation even though they are preferentially situated at high redshifts, and most of the intensity attributed to FSRQs occurs at lower energies.  Hence, while the BL Lacs (including cascades) dominate the collective blazar intensity at the highest energies, the FSRQs dominate at the lowest energies. In Figure \ref{fig-FSRQBLLadded}, we have plotted the superposition of the collective intensity of the BL Lacs and FSRQs for the Gilmore and Stecker EBL models and for two different relative abundances of BL Lacs and FSRQs. For the EGRET EGRB data sets, either of the scenarios for the relative abundances plotted could reproduce well the shape\footnote{However, note that there are additional systematic uncertainties (not plotted here) that could be quite large, especially at the highest energies.}. For the case in which BL Lacs comprise a lower fraction of the blazar population, the absorption feature is more prominent, as expected given the previous result for the LDDE model (Figure \ref{fig-BlazarEGRBLDDE}).

It should be noted that the effects of secondary cascades have been neglected. Even though the VHE electrons that are produced in the pair production processes of VHE photons can certainly produce VHE photons, the spectra of cascade photons above TeV energies are very steep.  Since most of the cascade radiation originated from the interactions of TeV photons with the infrared part of the EBL, the steep decline of such primary photons in the cascade spectrum ensures that the effect of secondary cascades is small. We have also neglected the possible contribution from synchrotron radiation resulting from the propagation of VHE electrons through intervening magnetic fields, which we will examine in a future publication.

\section{Discussion and Conclusions}
\label{cha:conclude}

We have demonstrated the effects of EM cascade radiation on the contribution to the EGRB attributable to blazar spectra finding that the cascade radiation \emph{flattens} the spectrum of the overall collective intensity of blazars and can greatly enhance their resulting contribution to the EGRB at the high-end of the \emph{Fermi} energy range. We have also demonstrated that the amount of cascade radiation is sensitive to the EBL model and the nature of blazar spectra. We make use of a distribution of blazar gamma-ray spectral indices as determined from the most recent data provided by \emph{Fermi} taking into account the possible spread in the distribution. Such a spread in the SID results in curvature in the collective intensity, which impacts the number of VHE photons available to create cascades. Furthermore, we have applied a similar analysis to two blazar sub-populations, BL Lacs and FSRQs, distinguishing between their spectral characteristics and their evolution with cosmic time. We found that while the inclusion of cascade radiation does not alter much the spectrum of the collective intensity of FSRQs, it can considerably enhance the collective intensity of BL Lacs. Hence, the FSRQs will dominate the collective intensity at lower energies and the BL Lacs at higher energies. In comparison to the aforementioned analyses (see Section \ref{cha:intro}) that include semiempirical determinations of the cascade radiation, the spectrum of this analysis appears different owing to the inclusion of the effect of the spread in the blazar spectral index and the distinct considerations of the evolution and spectral characteristics of the BL Lacs and FSRQs. In fact, the distinguishing features of this analysis allow the unique outcome of producing a collective blazar spectrum that reproduces well the shape of the spectrum of the EGRET EGRB. 

As demonstrated in this paper, EM cascades can greatly enhance the contribution to the EGRB attributable to blazars, particularly if the spread in the blazar SID is large or consists of relatively hard mean spectral indices and blazar spectra do not change appreciably at VHEs. As the cascade radiation, as well as the emission from the intrinsic spectra, is subject to absorption, the overall collective blazar intensity will exhibit an absorption feature at the highest energies, the shape of which is sensitive to the blazar GLF and the model of the EBL. The enhancement of the high-energy intensity by cascade radiation allows the high-energy absorption feature to be more easily measured by \emph{Fermi}. On the other hand, in the case that BL Lacs and FSRQs form distinct blazar sub-populations with respect to spectral characteristics and evolution \citep[as indicated by][]{lat09a}, the cascade radiation from the BL Lacs could account for a significant fraction of the BL Lac contribution to the collective blazar intensity, particularly at high energies. If the collective intensity from BL Lacs is concentrated at low redshifts, then even though their cascade radiation may be considerable, their collective spectrum above $10$ GeV would exhibit little attenuation. Hence, a large fraction of BL Lacs in the blazar population as compared to the FSRQs could result in a collective spectrum for which the absorption feature is less prominent. Thus, the observation of the absorption feature by \emph{Fermi} would provide a unique probe into the evolution of blazars with cosmic time as well as the nature of the EBL.  In fact, as shown in the figures, the preliminary \emph{Fermi} EGRB\footnote{M. Ackermann, TeV Particle Astrophysics 2009} is intriguingly steep.  However, as this is only a preliminary measurement (and without luminosity functions based on \emph{Fermi} blazars and having calculated for the EGRET sensitivity), it is, as yet, premature to draw conclusions from comparisons with this analysis. Nonetheless, we will return to this question in the near future.

As noted in Section \ref{subsec-blsp}, it is possible that over the course of so many decades in energy, blazar spectra could change, most likely turning over at some energy. In such a case, the cascade radiation could be reduced considerably and the collective intensity of blazars might not be appreciably enhanced at high energies. The most notable effect is that on the contribution from BL Lacs. The flattening of the spectrum at high energies is due, at least in part, to the enhanced cascade radiation. If BL Lac spectra break at energies low enough to render the cascade radiation insubstantial, the collective spectrum at high energies would be steeper. If the break occurs at energies greater than $100$ GeV, then the collective spectrum at high energies would still be more enhanced than it would be for the collective intensity from FSRQs alone and the absorption feature could be observable. If the break occurs at energies on the order of tens of GeV, then the collective spectrum would be steeper than in the either of the previous scenarios due to both the steeping of the intrinsic spectrum and absorption. If even without the enhancement from cascade radiation blazars comprise the bulk of the EGRB, the low-energy-break and high-energy-break scenarios would be distinguishable from each other, as well as distinguishable from the no break scenario.  Thus, the study of the collective intensity of blazars and the measurement of the EGRB by \emph{Fermi} will have implications for the nature of blazar spectra at VHEs, which would be \emph{complimentary} to the constraints provided by measurements of individual blazar spectra by IACTs \citep{alb08}.

Additionally, it is possible that blazar spectra change not only at high energies, but also during periods of increased activity \citep[see, e.g.,][]{von95,ss96a,bot99}; hence, the SIDs of such flaring blazars would be distinct from that of quiescent blazars, affecting their respective contributions to the collective blazar intensity. However, the question of whether blazar spectra change during flaring periods remains, as yet, unresolved due in large part to the lack of time resolution in EGRET data \citep{vp07}. With the improved sensitivity of \emph{Fermi}, time-resolved spectroscopy will be possible allowing the measurement of blazar spectra during flaring periods and the determination of whether flaring and quiescent blazars do indeed comprise spectrally distinct populations.

It should be noted that the question of blazar statistics and their role in determining the blazar contribution to the EGRB is complicated by the presence of intervening magnetic fields.  With negligible magnetic fields, the relativistic beaming of the intrinsic gamma-ray emission from AGNs ensures that observable gamma-rays originate only from those AGNs that are favorably aligned with respect to the observer (i.e., blazars).  This is true both for the intrinsic gamma-ray emission from AGNs and also for the cascade radiation, since without magnetic fields, the cascades are highly collimated in the direction of the propagating VHE photons.  In the presence of substantial magnetic fields, the cascades will no longer be collimated, increasing the number of gamma-ray emitting AGNs contributing to the EGRB. The lack of collimation of the cascades would also introduce halos of gamma-ray emission around blazars, which would contribute to source confusion.  On the other hand, the presence of substantial magnetic fields would introduce a synchrotron component in the cascades redistributing the energy in the spectrum of cascade radiation.  Though, the study of the effects of intervening magnetic fields would be insightful, their inclusion would further complicate this analysis.  Thus, the consideration of magnetic fields is out of the scope of this paper; however, we will return to this question in a future publication.

It is also possible that blazars cannot entirely account for the EGRB or could account for the EGRB only at lower energies. In such a case, the participation of other sources of high-energy gamma rays would be required to fully explain the EGRB, and the spectrum of the EGRB would also exhibit interesting features. For example, though blazars currently dominate the identifications of extragalactic TeV gamma-ray emitters, it is by no means clear that blazars are the only sources of TeV gamma rays.  If other astrophysical objects, such as GRBs and starburst galaxies, are found to also emit TeV gamma rays, they too will produce radiation from cascades. However, as likely candidate extragalactic TeV sources also constitute cosmological populations, they too will exhibit absorption at the highest energies. In addition to cascades from VHE photons, secondary particles produced through the propagation of ultra-high-energy cosmic rays through the EBL and the CMB will initiate cascades which will also contribute to the EGRB \citep{ste70,der86,lee98,ss99}. The origins of ultra-high-energy cosmic rays are, as yet, unknown; hence, the study of the EGRB together with \emph{Fermi} measurements can have profound implications for the highest energy cosmic rays.

On the other hand, if the collective intensity from some other gamma-ray source \emph{peaks} at high energies, then the resulting EGRB would also flatten due to the transition and could turn over depending on the spectrum of the overtaking gamma-ray source. In such a case, \emph{Fermi}-measured blazar GLFs and SIDs and information about the high-energy properties of other gamma-ray sources will be necessary to distinguish between a transition and a higher energy break in blazar spectra. An intriguing example to consider would be that of gamma-ray emission from dark matter annihilation, which would peak near the mass of the dark matter particle. If dark matter annihilation contributes substantially to the EGRB (even if only at energies near the mass of the dark matter particle), then it will be revealed through anisotropy searches \citep[see, e.g.,][]{ando07,sie09}.

Thus, the study of the collective blazar intensity, including the effects of cascade radiation, combined with the measurements of the EGRB and the blazar gamma-ray luminosity-redshift distribution by \emph{Fermi} can provide a wealth of insight into the nature of blazar spectra, as well as the evolution of blazars and the EBL with cosmic time.

\acknowledgments{We acknowledge enlightening feedback from Angela Olinto, Vasiliki Pavlidou, Kostas Tassis, Floyd Stecker, Joel Primack, and Marjorie Corcoran.
This work was supported in part by the Kavli Institute for Cosmological Physics at the University of Chicago through grants NSF PHY-0114422 and NSF PHY-0551142 and an endowment from the Kavli Foundation and its founder Fred Kavli. T.~M.~V. was also supported by the NSF Graduate Research Fellowship Program.}

\bibliography{vent_arxiv_bibtex}
\bibliographystyle{apj}

\appendix

\section{Derivation of the Spectrum of Electromagnetic Cascade Radiation}\label{app-detderiv}

To determine the spectrum of cascade photons, we begin with the spectrum of scattered radiation per electron:
\begin{equation}
\frac{dN_{\Gamma}}{dE} = \int \!\! \frac{d^3N_{\Gamma'}}{dtdEd\epsilon}dtd\epsilon = \int \!\! \frac{d^3N_{\Gamma'}}{dtdEd\epsilon}\left|\frac{dt}{dE_e}\right|dE_ed\epsilon\,,
\end{equation}
where $\Gamma$ is the initial Lorentz factor of the electron, $\Gamma'$ is the Lorentz factor of the electron at some later time $t$, and we are integrating over the soft photon background and the inverse Compton timescale, which is determined by the electron energy losses\footnote{At the energies that we are considering in this paper, the continuous energy loss description for electrons is accurate even at energies that are no longer quite in the Thomson regime.  However, in the extreme Klein--Nishina regime, electrons can lose large fractions of their energies in individual interactions rendering the continuous energy loss approximation invalid.}, $|dt/dE_e|$. The differential scattered photon spectrum is determined by integrating the differential Klein--Nishina cross section \citep{blu70}:
\begin{equation}\label{eqn-diffscatspec}
\frac{d^3N_{\Gamma'}}{dtdEd\epsilon} = \frac{3\sigma_{T}c}{4\Gamma'^2}\frac{n(\epsilon)}{\epsilon}\left[2q\ln q+\left(1+2q\right)\left(1-q\right)+\frac{1}{2}\frac{(\Gamma_\epsilon q)^2}{1+\Gamma_\epsilon q}(1-q)\right]\,,
\end{equation}
where $n(\epsilon)d\epsilon$ is the soft photon background number density\footnote{Note that the soft photon background is redshift dependent.  However, the inverse Compton timescale is sufficiently short so that the soft photon background does not change appreciably during the creation of the cascade.}, $\Gamma_\epsilon = 4\epsilon \Gamma'/mc^2$, \linebreak
$q = \eta/\Gamma_\epsilon(1-\eta)$, $\eta = \epsilon_1/\Gamma' mc^2$, and $\epsilon_1$ is the scattered photon energy. The spectrum of scattered photons also defines the electron loss rate:
\begin{equation}
\frac{dE_e}{dt} = -\int \!\!\! \int (\epsilon_1-\epsilon)\frac{d^3N_{\Gamma'}}{dtd\epsilon d\epsilon_1}d\epsilon d\epsilon_1\,,
\end{equation}
since adiabatic losses due to cosmic expansion that occur between interaction points for the electron are negligible compared to the inverse Compton losses. The spectrum of inverse Compton scattered photons arising from the distribution of electrons per primary photon is determined by integrating over the spectrum of electrons, $dN_e = N_e(\Gamma)d\Gamma$:
\begin{equation}
\frac{dN}{dE} = \int \! \frac{dN_{\Gamma}}{dE}dN_e = \int \! \frac{dN_{\Gamma}}{dE}N_e(\Gamma)d\Gamma\,.
\end{equation}
We can determine the cascade spectrum by integrating over the spectrum of primary photons, $dN_\gamma = N_\gamma(E_p)dE_p$:
\begin{equation}
\frac{dN_{\rm tot}}{dE} = \int \frac{dN}{dE}dN_\gamma = \int \frac{dN}{dE} N_\gamma(E_p)dE_p.
\end{equation}
However, in the case of any photon interaction process that cannot be considered continuous, the ``spectrum'' of electrons per primary photon is discrete. In this case, the resulting spectrum of inverse Compton scattered photons is best described as
\begin{equation}
\frac{dN}{dE} = \sum_{j=1}^{N_e} \frac{dN_{\Gamma_j}}{dE}\,.
\end{equation}
For a discrete distribution of primary photons (i.e., Monte Carlo photon propagation), the total cascade spectrum is given by
\begin{equation}
\frac{dN_{\rm tot}}{dE} = \sum_{k=1}^{N_\gamma} \sum_{j=1}^{N_e^k} \frac{dN_{\Gamma_j^k}}{dE}\,,
\end{equation}
where $N_e^k$ is the number of electron-like particles produced in the interaction of the $k^{\rm th}$ primary photon, and $\Gamma_j^k$ is the Lorentz factor of the $j^{\rm th}$ electron resulting from the interaction of the $k^{\rm th}$ primary photon.  In the case of pair production, $N_e^k = 2$, $\Gamma_j^1 = f\times E_{p_j}/m_ec^2$, and $\Gamma_j^2 = (1-f)\times E_{p_j}/m_ec^2$, where $f$ is some fraction of the energy of the primary photon (determined from the kinematics of the interaction). The \emph{observed} spectrum from a particular epoch\footnote{Unlike electrons, photons can propagate a considerable distance before the point of interaction and adiabatic losses must be included. Hence, $E_p$ is the \emph{redshifted} energy of the primary photon.} would then be
\begin{equation}
\frac{dN_{\rm tot}}{dE_0}(z) = (1+z) \sum_{k=1}^{N_\gamma(z)} \sum_{j=1}^{N_e^k} \frac{dN_{\Gamma_j^k}}{dE}\,.
\end{equation}
The total \emph{observed} spectrum is determined by summing over all of the epochs:
\begin{equation}
\frac{dN_{\rm tot}}{dE_0} = \sum_{l=1}^{N_{\rm zbins}} (1+z_l) \Delta z_l \sum_{k=1}^{N_\gamma(z_l)} \sum_{j=1}^{N_e^k}\frac{dN_{\Gamma_j^k}}{dE}\,,
\end{equation}
where $N_{\rm zbins}$ is the number of redshift bins, $z_l$ is the redshift of the $l^{\rm th}$ bin, and $\Delta z_l$ is the size of the $l^{\rm th}$ redshift bin.  For a continuous spectrum of photons undergoing pair production interactions, $d^2N_\gamma/dzdE_p$, the total \emph{observed} spectrum is given by
\begin{equation}
\frac{dN_{\rm tot}}{dE_0}(E_0) = \int \!\!\! \int (1+z) \frac{d^2N_\gamma}{dzdE_p} P(f;E_p,z) \left[\frac{dN_{\Gamma_1}(E_0(1+z))}{dE}+\frac{dN_{\Gamma_2}(E_0(1+z))}{dE}\right]dE_p dz\,,
\end{equation}
where $P(f;E_p,z)$ is the probability that the pair production interaction of a primary photon of energy $E_p$ will produce electron-type particles of energies $f\times E_p$ and $(1-f)\times E_p$.
Including the attenuation factor, $\exp[-\tau(E_0,z)]$, we finally arrive at the \emph{observed} differential cascade spectrum:
\begin{eqnarray}
\frac{dN^c}{dE_0}(E_0) &=& \!\!\!\!\! \int \!\!\! \int (1+z) \frac{d^2N_\gamma}{dzdE_p} P(f;E_p,z) \left[\frac{dN_{\Gamma_1}(E_0(1+z))}{dE}\right] e^{-\tau(E_0,z)} dE_p dz \nonumber \\
&+& \!\!\!\!\! \int \!\!\! \int (1+z) \frac{d^2N_\gamma}{dzdE_p} P(f;E_p,z) \left[\frac{dN_{\Gamma_2}(E_0(1+z))}{dE}\right] e^{-\tau(E_0,z)} dE_p dz. \nonumber \\
\end{eqnarray}

\end{document}